\documentstyle[preprint,aps,tabularx,epsfig]{revtex}

\begin{document}
\draft

\title{Periodic orbit analysis of an elastodynamic resonator using shape deformation}

\author{T. Neicu and A. Kudrolli}

\address{Department of Physics, Clark University, Worcester, MA 01610}
\date{\today} 

\maketitle

\begin{abstract}

We report the first definitive experimental observation of periodic orbits (POs) in the spectral properties of an elastodynamic system. The Fourier transform of the density of flexural modes show peaks that correspond to stable and unstable POs of a clover shaped quartz plate. We change the shape of the plate and find that the peaks corresponding to the POs that hit only the unperturbed sides are unchanged proving the correspondence. However, an exact match to the length of the main POs could be made only after a small rescaling of the experimental results. Statistical analysis of the level dynamics also shows the effect of the stable POs.

\end{abstract}

\pacs{PACS number(s): 05.45.+b,05.45.Mt,46.40.-f }

%\begin{multicols}{2}

Mechanical wave systems show localization and band gaps similar to solid state systems raising the possibility of applications~\cite{Espinosa:1998,Torres:1999}. An issue of current interest is if techniques developed in the analysis of quantum wave systems with chaotic classical counterparts can be applied to elastodynamic systems. The first experiments on elastodynamic systems were conducted with aluminum blocks and a statistical analysis was carried out~\cite{Weaver:1989,Ellegaard:1995}. Waves can propagate in transverse and longitudinal modes in a solid and the application of analysis techniques developed in the context of quantum systems is not straightforward. There are two main approaches to analyzing quantum wave systems. The first is a statistical approach where the properties of the eigenvalues can be compared to the universal predictions of random matrix theory (RMT)~\cite{Bohigas:1984,Mehta:1990}. The second semiclassical approach uses the periodic orbits (POs) in the system as a framework to understand the structure of the eigenvalues~\cite{Gutzwiller:1990}. In recent experiments it has been shown that the flexural modes of plates show statistical properties described by RMT~\cite{Bertelsen1:1999,Schaadt:1999,Bertelsen:1999}. However, the experimental observation of POs in elastodynamic systems has proven to be difficult. 

A semiclassical theory using the biharmonic equation for the flexural modes of a thin plate suggests that a close correspondence between the modes and ray orbits may be expected for the elastodynamic system because of the similarity of the governing wave equations~\cite{Bogomolny:1998}. However, experiments indicate that higher order corrections are required to capture the dispersion relation in plates at higher frequencies~\cite{Bertelsen:1999}. Therefore the biharmonic equation is not always a good approximation for the flexural modes of a plate in a regime where semiclassical analysis can be applied. We reported in an earlier study using a clover shaped plate that the statistical properties indeed show deviations from universality due to the presence of stable POs, but the result of the periodic orbit analysis was unclear~\cite{Neicu:2001}. Earlier studies with aluminum blocks have also yielded ambiguous results for the periodic orbits~\cite{delande}.

In this paper, we report the first detailed experimental study of the signatures of the periodic orbits in the spectral properties of an elastodynamical system. We perform measurements on a clover shaped quartz plate which has both stable and unstable POs (see Table I). Fourier transform is performed on the flexural density of modes using recently developed theoretical results for the dispersion relation~\cite{Bertelsen:1999}. To test the robustness of the peaks, we change one of the parameters which defines the shape of the plate in small steps that also alters some of the POs. The peaks corresponding to POs which do not hit the perturbed side are observed to remain unchanged. Furthermore, the peaks corresponding to the POs which hit the perturbed side are observed to change, proving the correspondence. An exact match to the length of the main POs could be made only after a small rescaling of the experimental results. We also analyze the statistical properties of the elastodynamical resonances under parametric variation using the parametric number variance $v(x)$ and the velocity autocorrelation $c(x)$. 

In the experiments, a fused quartz plate of thickness $h = 1.5875$~mm with dimensions given in the caption to Table~I is used. The plate is kept on three piezoelectric transducers (one transmitter and two receivers) and the vibrations of the plate are measured using a HP4395A Network Analyzer. Therefore the plate vibrates freely inside the chamber, the only contacts are through tiny ruby spheres that are attached to the transducers which do not perturb the resonances. The plate and the transducers are placed inside a temperature controlled chamber at 300\,K and at a pressure below $10^{-1}$ Torr to prevent losses due to air damping. A more detailed description of the experimental setup can be found in  Ref.~\cite{Schaadt:1999}.

The vibrational modes of a thin plate can be divided into flexural and in-plane modes. These two types of modes obey different dispersion relations and it is therefore important to separate them before analyzing the data. To experimentally isolate the flexural class of modes, we increase the pressure inside the chamber, and measure the resonance spectrum at $\sim$ 300 Torr. The $Q$-factor of the flexural modes which is about $10^5$ decreases, whereas the $Q$-factor for the in-plane modes are unchanged~\cite{Schaadt:1999,Neicu:2001,Schaadt:thesis}. Thus the resonances corresponding to the flexural modes are identified.  

In the first series of experiments, we sanded off material at one of the edges of the plate [indicated by thick line in the inset to Fig.~\ref{Fig1}(a)] in 61 small steps. This causes PO number 3 (shown in Table I) to decrease in length while PO number 4 remains unchanged. The flexural modes were measured for each step in a frequency interval between 52 kHz (mode number 80) and 352 kHz (mode number 605) with a frequency resolution of 5/8 Hz. We note that all modes in this frequency regime are two dimensional as the wavelength is much greater than the thickness of the plate. In the second series of experiments, the same edge is first sanded at an angle to completely destroy PO number 3 and the resulting shape is shown in the inset to Fig.~\ref{Fig1}(b). Then the same edge is sanded in 60 small steps and the flexural modes in the frequency interval between 52 kHz (mode number 78) and 252 kHz (mode number 398) are measured. 

The change in the flexural modes as a function of mass removed $m$ due to sanding is shown in Fig.~\ref{Fig1} for the two series of experiments (the total mass of the unperturbed plate was 21.35140~g). We observe that a number of modes are observed to split in Fig.~\ref{Fig1}(a) because the C$_{4v}$ point group symmetry of the shape of the plate is broken. One may expect some modes that are quantized about unperturbed PO number 4 to remain unchanged. However, none of the modes are observed to remain constant. 

According to theoretical expectations, $|F(l)|^2$ versus length $l$ should show strong peaks at values corresponding to the length of the POs of the ray system which are listed in Table~\ref{PO}. Therefore to check for the signatures of the POs in the elastodynamical spectra, we calculate the square of the Gaussian-weighted Fourier transform for all the 121 sets of flexural modes using
\begin{equation}
|F(l)|^2 = \left|4 \sqrt\frac{\beta}{\pi}\int_0^{k_{max}} \frac{e^{-\beta k^2}}{k^r}e^{-ikl}\rho(k)dk \right|^2 \;\;\;,
\label{fft3_eq}
\end{equation}
where $\rho(k)$ represents the density of modes $\sum_{i\leq n}\delta(k-k_{i})$, $n$ is the total number of modes in a spectrum, and $k_{i}$ denotes the wavenumber of the $i$th mode. The parameter $\beta$ is chosen such as $e^{-\beta k_{max}^2} = 0.01$, a value sufficiently small to reduce the spurious oscillations of the peaks due to finite size effects, and $r = 0$ following Ref.~\cite{Bogomolny:1998}.  

A new dispersion relation that relates the wavenumber $k$ to the frequency $f$ for plates has been calculated in Ref.~\cite{Bertelsen:1999} and is given by:
\begin{equation}
k = {12^{1/4} \over {h \sqrt{\kappa}}} \sqrt{\Omega}(1+a_1\Omega+a_2\Omega^2),
\label{exact_disp_rel}
\end{equation}
where $\kappa=\sqrt{2/(1-\nu)}$, $\Omega=2\pi fh/c_s$, and $\nu$ is the Poisson's ratio. For fused quartz, $\nu = 0.16$, $c_s = 3750$~ms$^{-1}$, $a_1=0.177$, and $a_2=-9.40 \times 10^{-3}$. Until these recent results, only the first term in the expansion given in Eq.~\ref{exact_disp_rel} was used to calculate $k$.  

The $|F(l)|^2$ was obtained for each of the 121 sets of data and is plotted against $0.92 \times l$ in Fig.~\ref{Fig2}. A number of strong peaks are observed. The locations and the strength of some of the peaks change while others remain constant. As noted earlier, the first series of perturbation decreases the length of the PO number 3 while keeping the PO number 4 constant. Indeed we observe in Fig.~\ref{Fig2}(a) that the peak at the length corresponding to these orbits splits with one peak shifting to a lower value and the other remaining unchanged (see Inset to Fig.~\ref{Fig3} for detail). After the plate is perturbed to destroy PO number 3, the corresponding peak in Fig.~\ref{Fig2}(b) is observed to disappear.

We note that even the amplitude of the peaks changes in a way which is in agreement with the idea that the area of the tori associated with that PO contributes to the amplitude of the peak. For example the amplitude of the peak at 20.3 cm is decreasing as the plate is sanded and one of the tori is slowly destroyed (see Fig.~\ref{Fig3}). The amplitude is constant in the second series of experiments after the PO number 3 is completely destroyed (see Fig.~\ref{Fig2}(b)). 

However while making a comparison with the actual length of the POs, we find that the peaks in the $|F(l)|^2$ occur at a slightly higher value. This is the reason for the factor $0.92$ which multiplies the length $l$ while presenting the results in Fig.~\ref{Fig2}. The source of this factor is unclear and does not appear to correspond to errors in the values of $\nu$ and $c_s$. After the scaling, several other peaks can be matched with the POs listed in Table~\ref{PO}. Furthermore, it can be noted that the peaks corresponding to the POs which hit only the unperturbed dimensions are observed to be unchanged, while others are observed to change. 

We note that because the peaks are somewhat shifted our earlier study in which we examined only the $|F(l)|^2$ for one set of data was inconclusive. By parametrically varying the modes and examining the Fourier transform of each set of modes, we are able to now say with confidence that the peaks in the $|F(l)|^2$ correspond to the POs of the elastodynamic system. 

We also note that if the $k$ is calculated using only the first term in the dispersion relation given in Eq.~\ref{exact_disp_rel}, then the correspondence is not observed. Therefore the corrections to the dispersion relation which takes into account the three dimensional nature of the plate are necessary to make the correspondence. It is possible that an addition correction in the theoretical formula is required to account for the slight scaling which is still required to get an exact match.

We further analyze the statistical properties of the eigenvalues under parametric variation using the parametric number variance $v(x)$ and the velocity autocorrelation $c(x)$.  Deviations may be expected from the universal forms for chaotic systems calculated using the 0-d nonlinear $\sigma$-model~\cite{Simons:1993} because of the presence of stable POs. Most systems in nature have both stable and unstable POs, and therefore it is interesting to know the nature of the deviations. Therefore we also investigate the properties of the eigenvalues under parametric variation. The phase space explored by ray trajectories inside the clover geometry has been discussed in detail in Ref.~\cite{Olivier:2001}. The clover geometry has both chaotic and integrable regions in its phase space and it is therefore an example of a mixed system. To analyze the data, we normalize the energy $E$ corresponding to the flexural modes in units of the local mean level spacing $\Delta$ to obtain the normalized energy $\varepsilon = E/\Delta$ and the parameter $m$ in units of the square root of the local mean squared slope to give the rescaled external parameter 
$x = \sqrt{\left< \left({d\varepsilon \over dm}\right)^2 \right>} m$.

The parametric number variance $v(x)$ which measures the collective motion of levels under parametric variation ~\cite{Goldberg:1991} is defined as
$ v(x) = \langle (n(\varepsilon,x')-n(\varepsilon,x'+x))^2 \rangle \;\;,$ 
where $n(\varepsilon,x)$ is the staircase
function which  counts the number of energy levels at fixed $x$
with energy lower than $\varepsilon$, the average is over the normalized parameter $x'$ and $\varepsilon$. $v(x)$ for the clover shaped plate with C$_{4v}$ point group symmetry and the plate without any symmetries is plotted in Fig.~\ref{Fig4}(a). $v(x)$ for the plate without symmetries is observed to be greater than the universal form for a completely chaotic system~\cite{Simons:1993}. Greater deviations are present in case of the data corresponding the first series of experiments, where the symmetry of the plate is being broken and degeneracies lifted.  

To study the oscillations of the eigenvalues during perturbation, we calculated the velocity autocorrelation $c(x) = \left < {\partial \varepsilon(x') \over \partial x'}
{\partial \varepsilon(x'+x) \over \partial x'} \right >$, where the average is over the parameter $x'$ and the energy $\varepsilon$. For this correlator no analytical results exist for the intermediate values of $x$ therefore we compare our results to a curve numerically calculated by Mucciolo~\cite{Mucciolo:1996} which agrees with the analytical results in the
limit of large and small $x$~\cite{Simons:1993}. This form also agrees with the correlations observed in the spectra of plates with shapes that are completely chaotic~\cite{Schaadt:1999}. The result for $c(x)$ for the two series of experiments is shown in Fig.~\ref{Fig4}(b). The curve corresponding to the clover shaped plate without symmetries deviates from the universal results similar to that of $v(x)$. Therefore the presence of stable POs makes the oscillations slightly more irregular compared to the completely chaotic case which leads to smaller anti-correlations. Because of the added effect of breaking symmetry, the deviations from the universal results are even greater in the first series corresponding to Fig.~\ref{Fig1}(a).

In summary, the study provides the first conclusive evidence of the relevance of the periodic orbits in the analysis of elastodynamic systems. The ability to make this correspondence was possible only because of the systematic changes in the peaks of the Fourier transform due to shape deformation. However, even after using the improved relation for the dispersion relation, small discrepancy between the actual location of the peaks and the length of the periodic orbits is observed. We have further shown that the presence of the stable periodic orbits causes the level motions to fluctuate differently than the universal forms for completely chaotic systems. 

We thank Olivier Brodier, Kristian Schaadt, and Taimur Ellahi for assistance in setting up the experiments, and Joel Norton for technical assistance. This work was supported by Research Corporation. A.K. was partially supported by a Alfred P. Sloan Fellowship.

%\end{multicols}
%\newpage

\begin{table}
%\begin{center}
%\begin{tabular}{|c c c|c c c|} \hline
%Number & PO & $l_{po}$ & Number & PO & $l_{po}$\\
%   &         & (cm)   &    &         & (cm)  \\

%\hline
% 1 & \epsfig{file=traj7.eps,width=1cm}   & 14.5 & 7  & \epsfig{file=traj2_2.eps,width=1cm}& 22.4 \\
% 2 & \epsfig{file=traj7_7.eps,width=1cm} & 16.7 & 8  & \epsfig{file=traj4_4.eps,width=1cm}& 23.4\\
% 3 & \epsfig{file=traj1.eps,width=1cm}   & 20.3 & 9  & \epsfig{file=traj5.eps,width=1cm} & 28.7 \\
% 4 & \epsfig{file=traj1_1.eps,width=1cm} & 20.3 & 10 & \epsfig{file=traj2.eps,width=1cm} & 31.2 \\
% 5 & \epsfig{file=traj8.eps,width=1cm}   & 20.5 & 11 & \epsfig{file=traj6.eps,width=1cm} & 38.9 \\
% 6 & \epsfig{file=traj3_3.eps,width=1cm} & 21.1 & 12 & \epsfig{file=traj4.eps,width=1cm} & 64.5 \\

%\hline
%\end{tabular}
%\end{center}
\caption{The periodic orbit (PO) and their length $l_{po}$ for the clover shaped plate. The radius of the convex side is 3.556 cm, the radius of the concave side is 5.080 cm, and the distance between the longitudinal concave sides is 10.160 cm. PO number 1, 2, 5, 6 and 8 are unstable and the rest are stable.}
\label{PO}

\end{table}

%\newpage

\begin{figure}
%\center{\epsfig{file=Fig1.eps,width=14cm}}
\caption{(a) Representative motion of the unfolded flexural modes of the clover shaped plate as a function of the removed mass $m$ obtained by sanding the edge of the plate indicated by the thicker line in the inset. The perturbation preserves the reflection symmetry about the vertical axis and decreases the length and stability of PO number 3. (b) The motion of the unfolded level after the reflection symmetry of the plate and PO number 3 is destroyed.}
\label{Fig1}
\end{figure}

%\newpage

\begin{figure}
%\center{\epsfig{file=Fig2.eps,width=13cm}}
\caption{(a) $|F(l)|^2$ versus length $l$ rescaled by a factor of 0.92 for the first experimental series. $|F(l)|^2$ is obtained using approximately 500 flexural modes in the frequency interval 52 kHz to 352 kHz. (b) $|F(l)|^2$ versus length $l$ rescaled by a factor of 0.92 for the second series.  $|F(l)|^2$ is obtained in this case using 306 flexural modes in the frequency interval 52 kHz to 252 kHz. The position of the POs of the system are also indicated. The numbers corresponds to the POs shown in the Table.~\ref{PO}.}
\label{Fig2}
\end{figure}

%\newpage

\begin{figure}
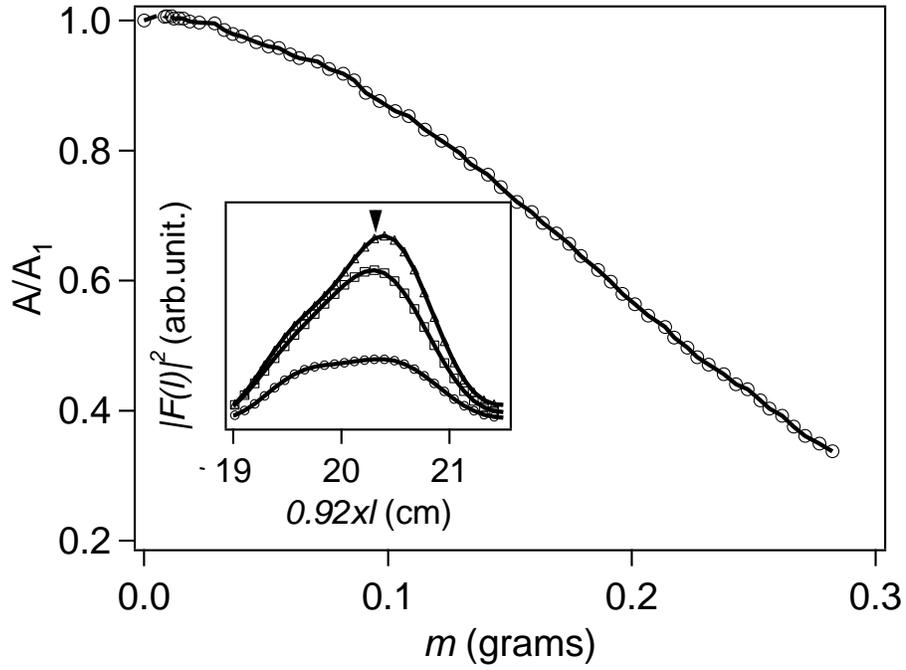

%\center{\epsfig{file=Fig3.eps,width=13cm}}
\caption{The ratio of the amplitude $A$ of the peak near PO number 4 normalized by the amplitude of peak $A_1$ corresponding to the 1st data set. Inset: A detailed plot of the peaks in $|F(l)|^2$ of the 1st, 31st and 61st data set in the parameter $m$ that correspond to PO number 3 and 4.}
\label{Fig3}
\end{figure}

%\newpage

\begin{figure}
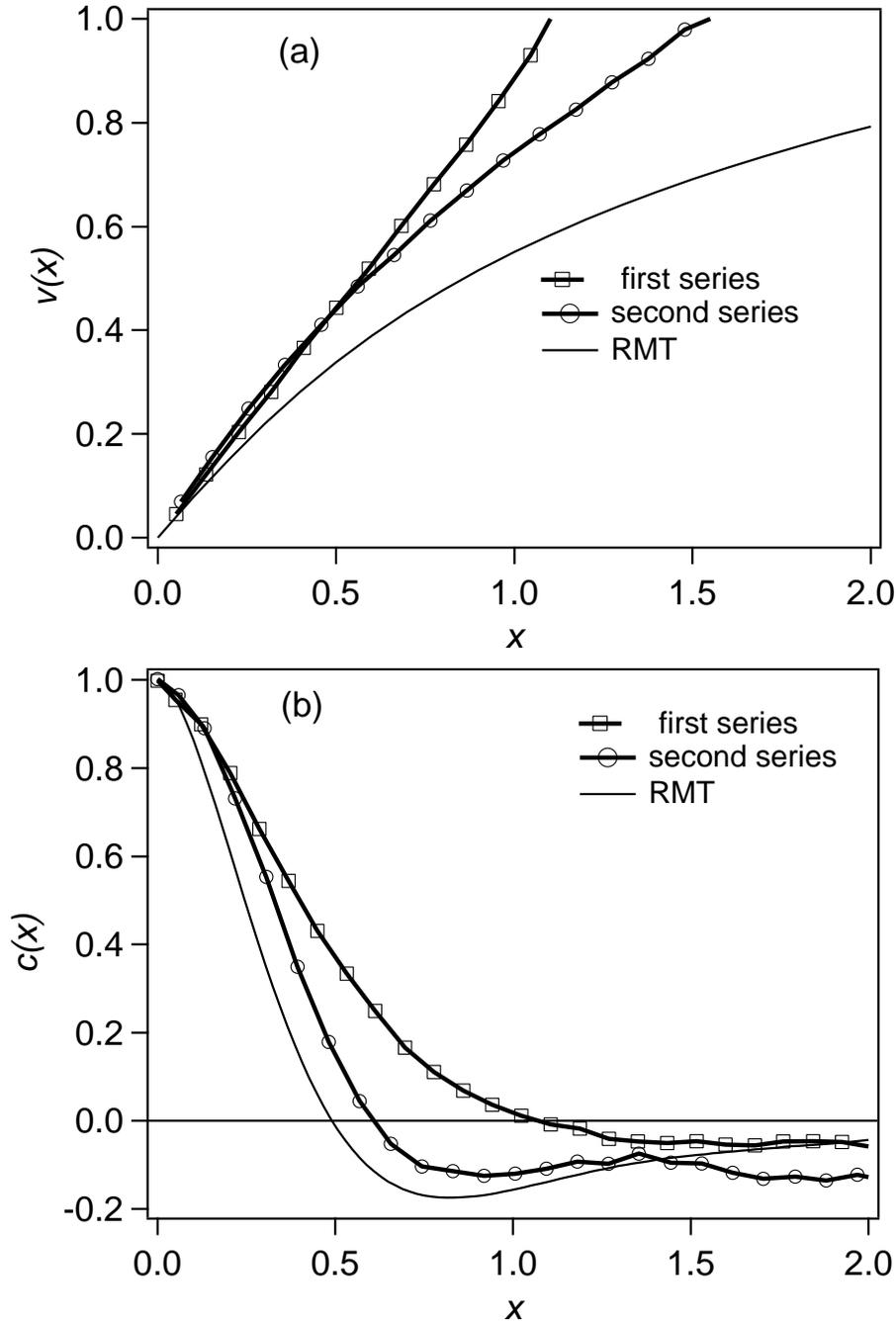

%\center{\epsfig{file=Fig4.eps,width=13cm}}
\caption{(a) The parametric number variance $v(x)$ versus the normalized parametric variation $x$ due to sanded mass $m$. (b) The velocity autocorrelation $c(x)$ versus $x$ for the experimental data compared to universal RMT curves. The deviations in the case of the second series is due to the presence of stable POs. The deviations are stronger in the first data series because the symmetry of the plate is being reduced.}
\label{Fig4}
\end{figure}

%\end{multicols}

\end{document}